\documentclass[conference]{IEEEtran}
\IEEEoverridecommandlockouts

\usepackage{cite}
\usepackage{amsmath,amssymb,amsfonts}
\usepackage{algorithm}
\usepackage{algpseudocode}
\usepackage{graphicx}
\usepackage{textcomp}
\usepackage{xcolor}
\usepackage{subcaption}

\usepackage{amsthm} 

\theoremstyle{definition}

\theoremstyle{remark}

\def\BibTeX{{\rm B\kern-.05em{\sc i\kern-.025em b}\kern-.08em
    T\kern-.1667em\lower.7ex\hbox{E}\kern-.125emX}}

\renewcommand{\baselinestretch}{0.84}
    
\begin{document}

\title{On the Optimality of Network Topology Discovery in Single-Hop Bounded-Interference Networks\\
}

\author{\IEEEauthorblockN{Tolunay Seyfi\(^{1}\), Erfan Khadem\(^{2}\), Fatemeh Afghah\(^{1}\)}
\IEEEauthorblockA{\(^{1}\)Holcombe Department of Electrical and Computer Engineering, Clemson University, Clemson, SC, USA}
\IEEEauthorblockA{\(^{2}\)Department of Electrical Engineering, Sharif University of Technology, Tehran, Iran}
\IEEEauthorblockA{tseyfi@clemson.edu, erfan.khadem@ee.sharif.edu, fafghah@clemson.edu}
}

\maketitle

\begin{abstract}
We propose \emph{PRISM} (\textbf{Pseudorandom Residue-based Indexed Scheduling Method}), a deterministic topology-discovery framework for single-hop wireless networks with bounded interference. Each receiver has at most \(L\) interfering transmitters among \(K\) transmitters and identifies them through singleton transmissions. PRISM assigns finite-field labels to transmitters and schedules transmissions via modular multiplication and a second prime modulus. It achieves full discovery in \(O(L(1+\delta)\log K)\) rounds in expectation with failure probability \(K^{-\delta}\), and in \(O(L^2\log K)\) rounds deterministically. Simulations show \(\approx 0.9L\log K\) scaling, with \(q/L\approx1.2\) minimizing mean completion time and \(q/L\approx1.4\text{--}1.6\) improving tail performance.
\end{abstract}


\section{Introduction}

Large-scale wireless systems require scalable and reliable mechanisms for topology discovery, i.e., for identifying which transmitters interfere with which receivers, since such information underlies scheduling, routing, interference management, and resource allocation~\cite{veeravalli2018interference}. Classical neighbor-discovery methods based on random access or probabilistic contention~\cite{vasudevan2013efficient} are often ill-suited to ultra-reliable and low-latency settings such as industrial wireless control~\cite{ahmed2005topology}, dense vehicular systems, and UAV networks. This has motivated heuristic approaches~\cite{bejerano2015fast,li2018hybrid}, learning-based methods~\cite{yang2017learning}, and structured code constructions such as on-off necklace codes for asynchronous discovery~\cite{tirkkonen2017onoff}. At the same time, lower bounds for radio communication highlight the intrinsic difficulty of fast deterministic discovery in adversarial settings~\cite{alon1989complexity}. Motivated by these limitations, we study deterministic, non-adaptive topology discovery in single-hop bounded-interference networks. Prior work~\cite{Seyfi2021} introduced a deterministic prime-indexed scheduling framework with round complexity scaling as $\log^2 K$ due to repeated prime reuse across phases. In contrast, we develop a multiplicative residue-based construction that uses a cyclic permutation over a finite field together with a second prime modulus to induce analyzable transmission patterns, thereby reducing systematic collisions without requiring feedback.

Our model is a bipartite interference graph $G=(T\cup R,E)$, where $T=\{T_1,\dots,T_K\}$ is the set of transmitters, $R=\{R_1,\dots,R_K\}$ is the set of receivers, and $(T_i,R_j)\in E$ indicates that transmitter $T_i$ is in the interference neighborhood of receiver $R_j$. For each receiver $R_j$, the neighborhood $\mathcal{N}(R_j)=\{i:(T_i,R_j)\in E\}$ satisfies $|\mathcal{N}(R_j)|\leq L$. Each edge \((T_i,R_j)\in E\) represents a potential interfering transmission from \(T_i\) at receiver \(R_j\), and the discovery goal is for each receiver to identify all transmitters in its interference neighborhood. This abstraction captures both structured locally connected interference patterns arising in sectorized cellular layouts and more general non-local interference topologies induced by beamforming, reuse, blockage, mobility, and programmable wireless environments~\cite{Annapureddy12,elgamal2020joint,Karacora17}. 


\subsection{Contributions}
This work makes the following contributions:
\begin{itemize}
    \item We propose PRISM, a deterministic scheduling framework based on modular residue classes that enables topology discovery in bounded-interference bipartite networks without adaptive feedback.

    \item We establish provable complexity guarantees: $\mathcal{O}(L(1+\delta)\log K)$ rounds in expectation over random topology realizations, and $\mathcal{O}(L^2\log K)$ rounds in the worst case.

    \item We show that PRISM is zero-error and fully deterministic, and we validate its practical performance through Monte Carlo simulations, including parameter tuning with respect to $q/L$, empirical scaling close to $0.9L\log K$, and a linear-time residue-alignment verification procedure for certifying deterministic convergence.
\end{itemize}

\section{Algorithm: Pseudorandom Residue-based Indexed Scheduling Method (PRISM)}
\label{sec1}


\subsection{Algorithm Description}

Choose a prime $p>K$, assign each transmitter $i$ a unique initial label $x_{i,0}\in\mathbb{Z}_p^*$, select a generator $g$ of $\mathbb{Z}_p^*$, and choose a second prime $q=O(L)$ with $q>L$, where $L$ is the maximum receiver degree. The modulus $q$ determines the number of rounds per phase and controls the collision rate.

The protocol proceeds in phases. In phase $\phi\ge 1$, transmitter $i$ updates its label as
\[
x_{i,\phi}=(g x_{i,\phi-1})\bmod p=(x_{i,0}g^\phi)\bmod p,
\]
and maps its current label to round
\[
y_{i,\phi}=x_{i,\phi}\bmod q.
\]
Thus phase $\phi$ consists of $q$ rounds indexed by $j\in\{0,\dots,q-1\}$, and transmitter $i$ sends its original label $x_{i,0}$ only in round $j=y_{i,\phi}$. Each receiver starts with candidate set $\{1,\dots,K\}$ and refines it phase by phase. In round $j$: (i) if the round is silent, all candidates mapping to $j$ are eliminated; (ii) if exactly one transmitter $x_A$ is observed, then $x_A$ is declared a true neighbor and all candidates mapping to $j$ are removed; and (iii) if multiple transmitters collide, then the receiver cannot distinguish them and all candidates mapping to $j$ are retained. Hence a false candidate survives a phase only if it repeatedly falls into collision rounds. Set
\[
q=cL,\qquad c>1.
\]
As shown later,
\[
P_{\mathrm{survive}}
\le \frac{L^2}{2q^2}
= \frac{L^2}{2c^2L^2}
= \frac{1}{2c^2}.
\]
For example, $c=2$ gives $P_{\mathrm{survive}}\le 1/8$. Thus $c$ controls the trade-off between shorter phases and lower collision persistence. In practice, values near $c\approx 1.2$ minimize the mean completion time, values around $1.4$--$1.6$ improve tail latency, and larger values such as $c\in[2,3]$ are useful when seeking stronger deterministic separation guarantees. Let $M$ denote the number of phases. For a false candidate,
\begin{align}
\Pr[\text{survives }M\text{ phases}] &= (P_{\mathrm{survive}})^M, \\
K(P_{\mathrm{survive}})^M < K^{-\delta}
&\Longrightarrow
(P_{\mathrm{survive}})^M < K^{-(\delta+1)}, \\
M &> \frac{(\delta+1)\log K}{\log(1/P_{\mathrm{survive}})}.
\end{align}
Substituting $P_{\mathrm{survive}}=1/(2c^2)$ gives
\[
M=O((1+\delta)\log K).
\]
Since each phase contains $q=O(L)$ rounds, the total probabilistic runtime is
\[
Mq=O(L(1+\delta)\log K).
\]

\subsection{Deterministic Runtime Analysis}
\label{sec:deterministic}

We now show that even without probabilistic assumptions on the topology, a false candidate cannot remain indistinguishable from true neighbors indefinitely.

\subsubsection{Stuck Candidates Argument}
\label{sec1:Stuck_Candidates}

Call a false candidate $x_c$ \emph{stuck} if it collides with the same true neighbor over multiple consecutive phases. Let $x_{i,\phi}$ denote the index of a true neighbor $i$ in phase $\phi$, and let $x_{c,\phi}$ denote the corresponding index of candidate $c$. Suppose that in two consecutive phases $\phi-1$ and $\phi$,
\[
x_{c,\phi-1}\equiv x_{i,\phi-1}\pmod q,
\qquad
x_{c,\phi}\equiv x_{i,\phi}\pmod q.
\]
Using the update rule
\[
x_{i,\phi}=g x_{i,\phi-1}-\ell_i p,
\qquad
x_{c,\phi}=g x_{c,\phi-1}-\ell_c p,
\]
where
\[
\ell_i=\left\lfloor \frac{g x_{i,\phi-1}}{p}\right\rfloor,
\qquad
\ell_c=\left\lfloor \frac{g x_{c,\phi-1}}{p}\right\rfloor,
\]
we obtain
\begin{align}
g x_{i,\phi-1}-\ell_i p
&\equiv
g x_{c,\phi-1}-\ell_c p \pmod q, \\
g(x_{i,\phi-1}-x_{c,\phi-1})
&\equiv
(\ell_i-\ell_c)p \pmod q.
\end{align}
Since the nodes already collide in phase $\phi-1$,
\[
x_{i,\phi-1}\equiv x_{c,\phi-1}\pmod q,
\]
so the left-hand side is $0$ modulo $q$, implying
\[
(\ell_i-\ell_c)p\equiv 0\pmod q.
\]
Because $p$ and $q$ are coprime,
\[
\ell_i-\ell_c\equiv 0\pmod q.
\]
Now assume $g\le q$. Then
\[
0\le \ell_i,\ell_c<g\le q
\qquad\Rightarrow\qquad
|\ell_i-\ell_c|<q,
\]
so the congruence modulo $q$ forces
\[
\ell_i=\ell_c.
\]
Substituting back yields
\[
x_{i,\phi}-x_{c,\phi}=g(x_{i,\phi-1}-x_{c,\phi-1})\pmod p.
\]
Thus, as long as the candidate remains stuck to the same true neighbor, the difference is multiplied by $g$ at each phase. Unrolling over $N$ phases gives
\[
|x_{i,N}-x_{c,N}|=g^N|x_{i,0}-x_{c,0}|,
\]
ignoring modular wraparound to expose the raw growth. Although the actual dynamics are modulo $p$, the absolute difference cannot exceed $p$, so once
\[
g^N|x_{i,0}-x_{c,0}|>p,
\]
continued sticking becomes impossible. Hence the number of stuck phases is bounded by
\[
N\le 1+\log_g\!\left(\frac{p}{\Delta_0}\right),
\qquad
\Delta_0=|x_{i,0}-x_{c,0}|.
\]
Since stuck candidates must initially satisfy
\[
|x_{i,0}-x_{c,0}|=kq,\qquad k\in\mathbb{Z}^+,
\]
we obtain the conservative bound
\[
N\le 1+\log_g\!\left(\frac{p}{q}\right)
=1+\frac{\log(p/q)}{\log g}.
\]
Therefore, no false candidate can remain stuck to a single true neighbor for more than logarithmically many phases. In particular, this already implies $O(L\log K)$ discovery time for small values such as $L\le 3$.

\subsubsection{Window-Based Separation Analysis (General Case)}

For larger $L$, a false candidate may survive by colliding with different true neighbors across phases. To handle this case, we use a window-based covering argument over the multiplicative sequence induced by $g$.

\noindent\textbf{Phase behavior and collision model:}
In phase $\phi$,
\[
x_{i,\phi}=g^\phi x_{i,0}\bmod p.
\]
Two nodes $i$ and $j$ collide if
\[
x_{i,\phi}\equiv x_{j,\phi}\pmod q,
\]
equivalently,
\[
g^\phi(x_{i,0}-x_{j,0})\bmod p \equiv 0 \pmod q.
\]
Let
\[
r_{ij}=x_{i,0}-x_{j,0}.
\]
Then a collision occurs when
\[
g^\phi r_{ij}\bmod p \bmod q \in \{0,\,p\bmod q\},
\]
where the two cases correspond to the two possible orderings of
\[
g^\phi x_{i,0}\bmod p
\quad\text{and}\quad
g^\phi x_{j,0}\bmod p.
\]

\noindent\textbf{Boolean collision window:}
Consider the sequence
\[
\{g^\phi r_{ij}\bmod p\}_{\phi=0}^{W-1}
\]
and mark phase $\phi$ as \texttt{True} whenever
\[
g^\phi r_{ij}\bmod p \bmod q \in \{0,\,p\bmod q\}.
\]
Because multiplication by $g$ permutes $\mathbb{Z}_p^*$, this sequence behaves like a cyclic traversal of residue values, and changing the pair $(i,j)$ only cyclically shifts the same sequence. The number of residues in $\mathbb{Z}_p$ satisfying
\[
x\bmod q\in \{0,p\bmod q\}
\]
is approximately $(2/q)p$, so
\[
P[\texttt{True}]=\frac{2}{q}.
\]
This gives a certifiable sufficient condition for separation: for fixed $(p,q,g)$, one can compute
\[
\{g^\phi r\bmod p \bmod q\}
\]
and verify in linear time whether every window satisfies the required separation property. Since different pairs correspond only to cyclic shifts, it suffices to check one representative sequence. In our implementation, we perform this verification after parameter selection; if it fails, one may increase the window length $W$.

\noindent\textbf{Sliding window argument:}
Fix a window of $W$ phases and let $N_{\mathrm{hits}}$ be the number of \texttt{True} entries in that window. Then
\[
\mathbb{E}[N_{\mathrm{hits}}]=\frac{2W}{q}.
\]
Set
\[
\mu=\frac{2W}{q},
\qquad
t=2\mu=\frac{4W}{q}.
\]
Applying the Chernoff bound,
\begin{align}
\Pr[N_{\mathrm{hits}}\ge t]
&\le
\exp(-\mu)\left(\frac{e\mu}{t}\right)^t \\
&=
\exp\!\left(-\frac{2W}{q}\right)\left(\frac{e}{2}\right)^{\frac{4W}{q}}.
\end{align}
Hence the tail probability decays exponentially in $W/q$.

\noindent\textbf{Union bound over all shifts:}
There are at most $p$ cyclic shifts of the sequence, so
\[
\Pr[\exists \text{ window with } N_{\mathrm{hits}}>2\mathbb{E}]
\le
p\exp\!\left(-0.76\,\frac{W}{q}\right).
\]
To make this probability smaller than $1$, it is sufficient that
\[
p\,e^{-0.76W/q}<1,
\qquad\Rightarrow\qquad
W>1.33\,q\log p.
\]

\noindent\textbf{Separation guarantee:}
Take
\[
W=2q\log p.
\]
Then, with high probability, every window of length $W$ contains at most
\[
2\cdot \frac{2W}{q}
=
\frac{4W}{q}
=
O\!\left(\frac{W}{q}\right)
\]
marked collision positions. A false candidate can survive only by aligning with at least two true neighbors per phase, so over $W$ phases it would require at least $2W$ admissible collision slots across the $L$ relevant sequences. But the total number of marked positions available across those $L$ windows is at most
\[
L\cdot \frac{4W}{q}.
\]
Therefore survival is impossible whenever
\[
L\cdot \frac{4W}{q}<2W
\qquad\Longleftrightarrow\qquad
L<\frac{q}{2},
\]
which holds by construction when $q>2L$.

\noindent\textbf{Final deterministic bound:}
Thus no false candidate can survive beyond
\[
W=O(q\log p)
\]
phases. Since each phase consists of $q$ rounds, the total deterministic runtime is
\[
Wq=O(q^2\log p)=O(L^2\log K),
\]
using $q=O(L)$ and $\log p=O(\log K)$. This establishes deterministic convergence for arbitrary $(K,L)$-bounded interference topologies.

\subsection{Probabilistic Runtime Analysis}

We now derive the expected number of phases needed to eliminate all false candidates with high probability.

Let a receiver have $s\le L$ true neighbors, and let $x_C$ be a false candidate. In phase $\phi$, every node computes
\[
y_{i,\phi}=x_{i,\phi}\bmod q.
\]
Candidate $x_C$ survives phase $\phi$ only if its mapped round $y_{C,\phi}$ collides with at least two true neighbors; silence or a singleton eliminates it. Hence survival requires a triple collision. There are
\[
\binom{s}{2}=\frac{s(s-1)}{2}
\]
pairs of true neighbors, and each pair maps to the same round with probability approximately $1/q$. Therefore
\[
\mathbb{E}[N_{\mathrm{collisions}}]\le \frac{L(L-1)}{2q}.
\]
Assuming pessimistically that all such pairwise collisions occur in distinct rounds, we obtain
\[
P_{\mathrm{survive}}<\frac{L^2}{2q^2}.
\]
With
\[
q=cL,\qquad c>1,
\]
this becomes
\[
P_{\mathrm{survive}}<\frac{1}{2c^2}.
\]
If $M$ phases are used, then
\begin{align}
P_{\mathrm{total}} &= (P_{\mathrm{survive}})^M, \\
K(P_{\mathrm{survive}})^M < K^{-\delta}
&\Longrightarrow
(P_{\mathrm{survive}})^M < K^{-(\delta+1)}, \\
M &> \frac{(\delta+1)\log K}{\log(1/P_{\mathrm{survive}})}.
\end{align}
Each phase uses $q=O(L)$ rounds, so the total runtime is
\[
Mq
=
O\!\left(
L\cdot
\frac{(\delta+1)\log K}{\log(1/P_{\mathrm{survive}})}
\right)
=
O(L\log K)
\]
for constant $\delta$ and fixed $c>1$.

The analysis relies on the fact that $x_{i,\phi}\bmod q$ behaves approximately like a pseudorandom uniform mapping across phases, justified by the full-cycle permutation induced by multiplication by a primitive root $g$ in $\mathbb{Z}_p^*$.
\section{Performance Evaluation and Simulation Results}

We evaluate PRISM to validate the predicted scaling laws and quantify the effect of the design parameter \(q\) on discovery efficiency. Unless otherwise stated, simulations are conducted in a single-hop collision-channel setting with \(K\) transmitters and \(K\) receivers, where each receiver has an interference neighborhood of \(L\) transmitters drawn uniformly at random under the \((K,L)\)-bounded degree model. We sweep \(K\in[128,7234]\), use \(200\) topology realizations per configuration, and consider \(L\in\{3,\ldots,12\}\). All algorithms are evaluated under the same collision-channel abstraction, topology realizations, and completion criterion. Performance is measured in \emph{communication rounds}, i.e., the total number of rounds until all receivers complete topology discovery.

\begin{figure}[t]
\centering
\begin{subfigure}[t]{0.48\textwidth}
    \centering
    \includegraphics[width=\linewidth,height=0.28\textheight,keepaspectratio]{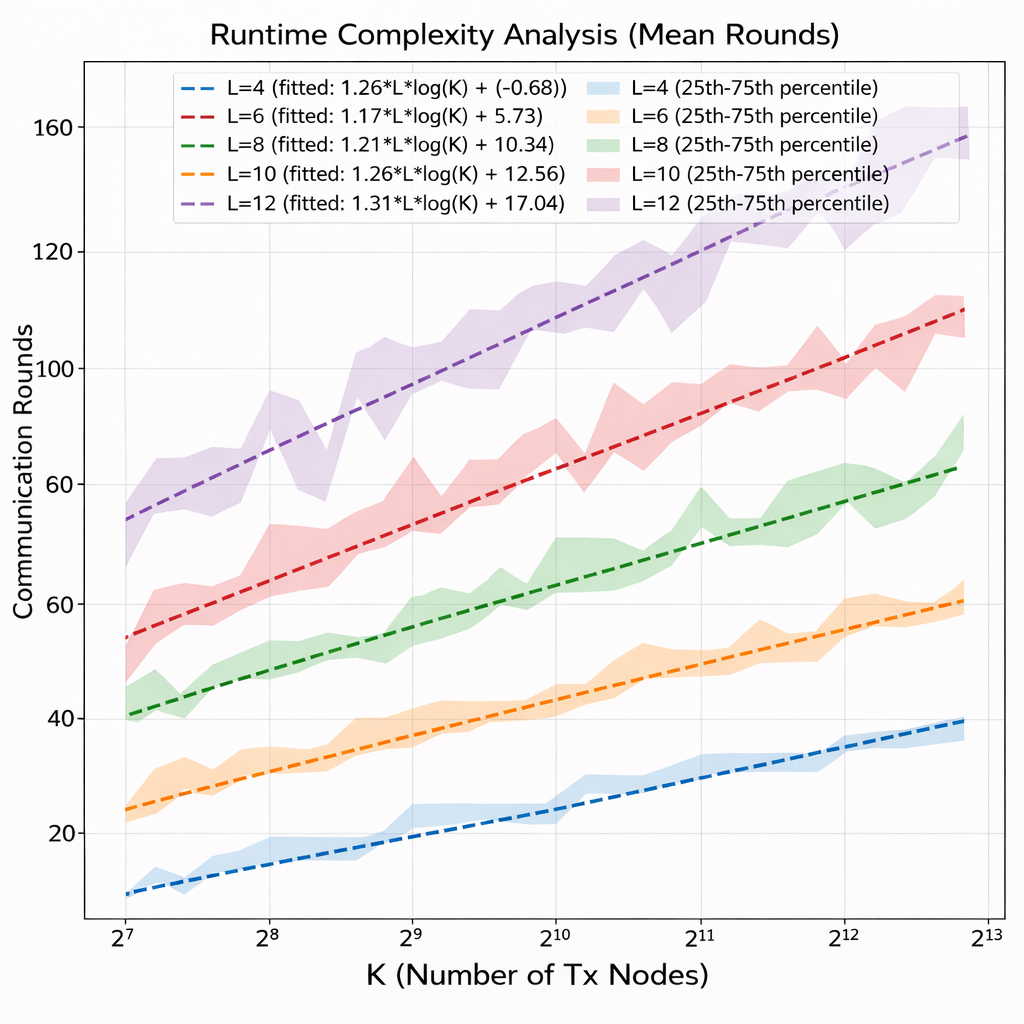}
    \caption{Mean and interquartile range of discovery rounds versus network size \(K\) at \(q/L=1.2\). The observed scaling is well approximated by \(0.9\,L\log K\), and the narrow interquartile range indicates low variance across topology realizations.}
    \label{fig:scaling_mean}
\end{subfigure}
\hfill
\begin{subfigure}[t]{0.48\textwidth}
    \centering
    \includegraphics[width=\linewidth,height=0.28\textheight,keepaspectratio]{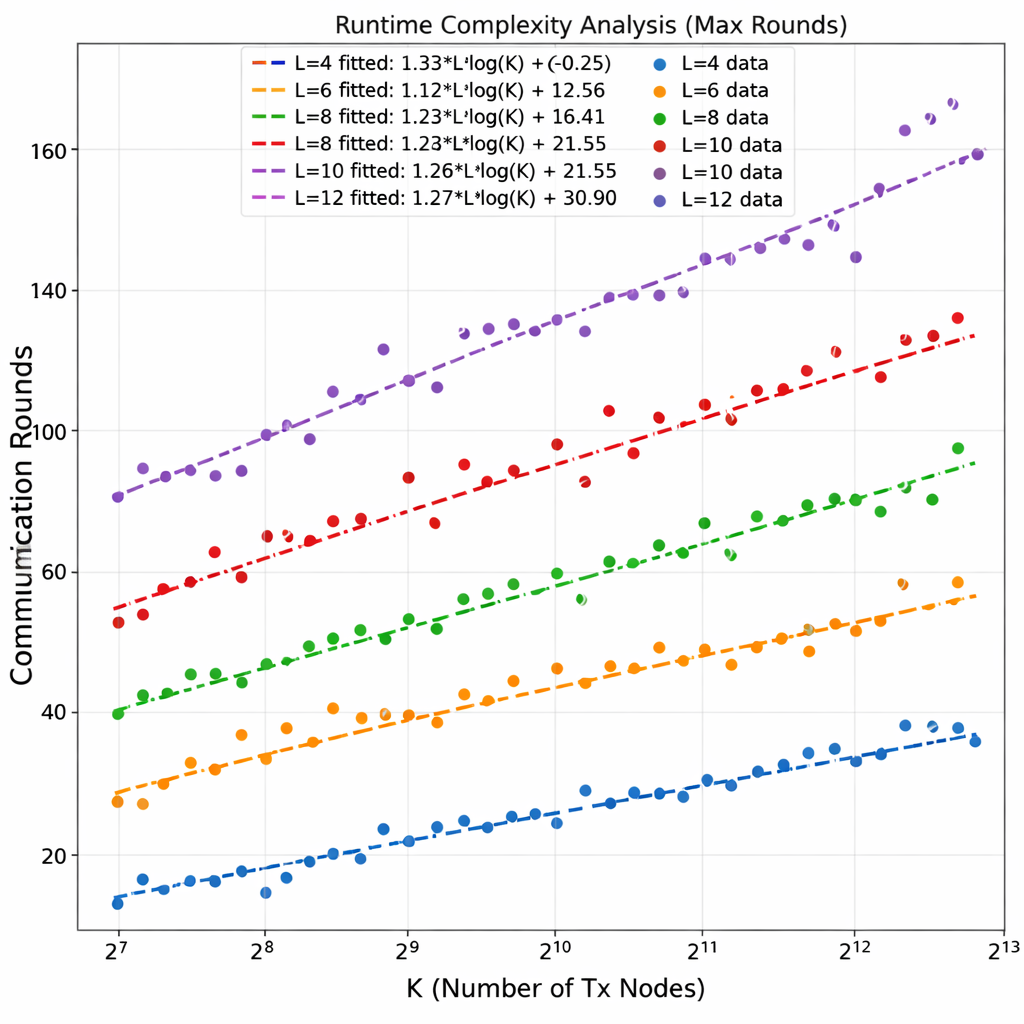}
    \caption{Maximum discovery rounds versus network size \(K\) at \(q/L=1.6\). The maximum completion time follows an approximate \(0.9\,L\log K + C\) trend while preserving logarithmic scaling.}
    \label{fig:scaling_max}
\end{subfigure}
\caption{Scaling behavior of PRISM.}
\label{fig:scaling_combined}
\end{figure}

Each round is treated as one abstract unit of control-plane effort, independent of physical-layer details such as modulation, coding, and bandwidth. We report two metrics: (i) the mean network completion time, defined as the average number of rounds required for the last receiver to finish across \(200\) realizations, and (ii) the maximum completion time across those realizations, capturing rare high-collision tail events. A key PRISM parameter is the per-phase modulus \(q\). The ratio \(q/L\) trades off phase length and collision density: increasing \(q\) improves transmitter separation but increases the number of rounds per phase, whereas decreasing \(q\) shortens each phase but increases persistent collisions. To identify effective operating points, we generate heatmaps versus \(K\) and \(q/L\), showing that the mean completion time is minimized near \(q/L\approx 1.2\), while the maximum completion time is minimized near \(q/L\approx 1.6\), indicating improved robustness to rare collision-heavy realizations. We simulate PRISM according to Algorithm~\ref{alg:dris}: each transmitter is assigned a unique identifier in \(\mathbb{Z}_p^*\) and transmits in exactly one round per phase according to the residue-cycling rule; each receiver monitors all \(q\) rounds and updates its candidate set based on silence, singleton, or collision outcomes. The protocol terminates when every receiver identifies its full neighborhood, and the same completion criterion is used for all baselines. Using these empirically selected operating points, we evaluate the scaling behavior of PRISM. Figure~\ref{fig:scaling_mean} shows the mean completion time versus \(K\) for \(q/L=1.2\), exhibiting a clear logarithmic dependence with fitted trend \(0.9\,L\log K\), consistent with the bound \(O(L\log K)\). Figure~\ref{fig:scaling_max} shows the corresponding maximum completion time for \(q/L=1.6\), which also follows an approximately logarithmic trend of the form \(0.9\,L\log K + C\). Across all configurations, the variance remains small relative to the mean, while the maximum metric captures rare but severe collision realizations. These results confirm that PRISM achieves predictable and scalable topology discovery under collision-channel constraints.

\begin{algorithm}[t]
\caption{Simulated Execution of PRISM}
\label{alg:dris}
\begin{algorithmic}[1]
\State \textbf{Input:} \(K,L,p,q,g\)
\State Assign each transmitter \(i\) a unique label \(x_{i,0}\in\mathbb{Z}_p^*\)
\State For each receiver \(j\), initialize \(\mathcal{C}_j\gets\{1,\dots,K\}\), \(\mathcal{N}_j\gets\emptyset\)
\For{each phase \(\phi=1,2,\ldots\)}
    \For{each transmitter \(i=1,\dots,K\)}
        \State \(x_{i,\phi}\gets (g x_{i,\phi-1})\bmod p\)
        \State \(y_{i,\phi}\gets x_{i,\phi}\bmod q\)
    \EndFor
    \For{each receiver \(j=1,\dots,K\)}
        \For{each round \(r=0,\dots,q-1\)}
            \State Observe active transmitters in round \(r\)
            \If{no transmission}
                \State \(\mathcal{C}_j\gets \mathcal{C}_j\setminus\{i\in\mathcal{C}_j: y_{i,\phi}=r\}\)
            \ElsIf{exactly one transmitter \(x_A\)}
                \State \(\mathcal{N}_j\gets \mathcal{N}_j\cup\{x_A\}\)
                \State \(\mathcal{C}_j\gets \mathcal{C}_j\setminus\{i\in\mathcal{C}_j: y_{i,\phi}=r\}\)
            \Else
                \State \textbf{continue}
            \EndIf
        \EndFor
        \If{\(\mathcal{C}_j=\emptyset\) \textbf{or} \(|\mathcal{N}_j|=L\)}
            \State Mark \(\mathrm{Rx}_j\) complete
        \EndIf
    \EndFor
    \If{all receivers are complete}
        \State \textbf{break}
    \EndIf
\EndFor
\State \textbf{Output:} Total rounds until full network discovery
\end{algorithmic}
\end{algorithm}

\begin{figure}[t]
\centering
\begin{subfigure}[t]{0.48\textwidth}
    \centering
    \includegraphics[width=\linewidth,height=0.28\textheight,keepaspectratio]{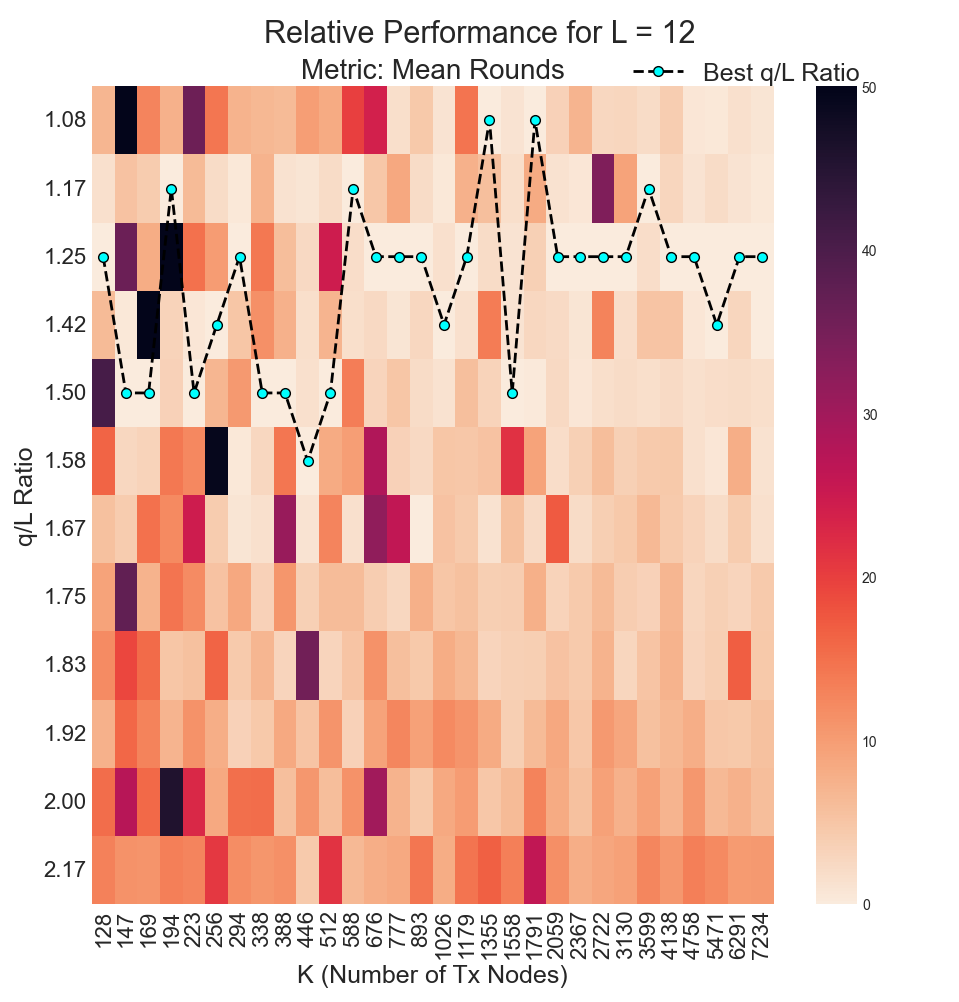}
    \caption{Relative degradation of the mean network completion time versus \(K\) and \(q/L\). The optimum occurs near \(q/L\approx 1.2\).}
    \label{fig:heatmap_both}
\end{subfigure}
\hfill
\begin{subfigure}[t]{0.48\textwidth}
    \centering
    \includegraphics[width=\linewidth,height=0.28\textheight,keepaspectratio]{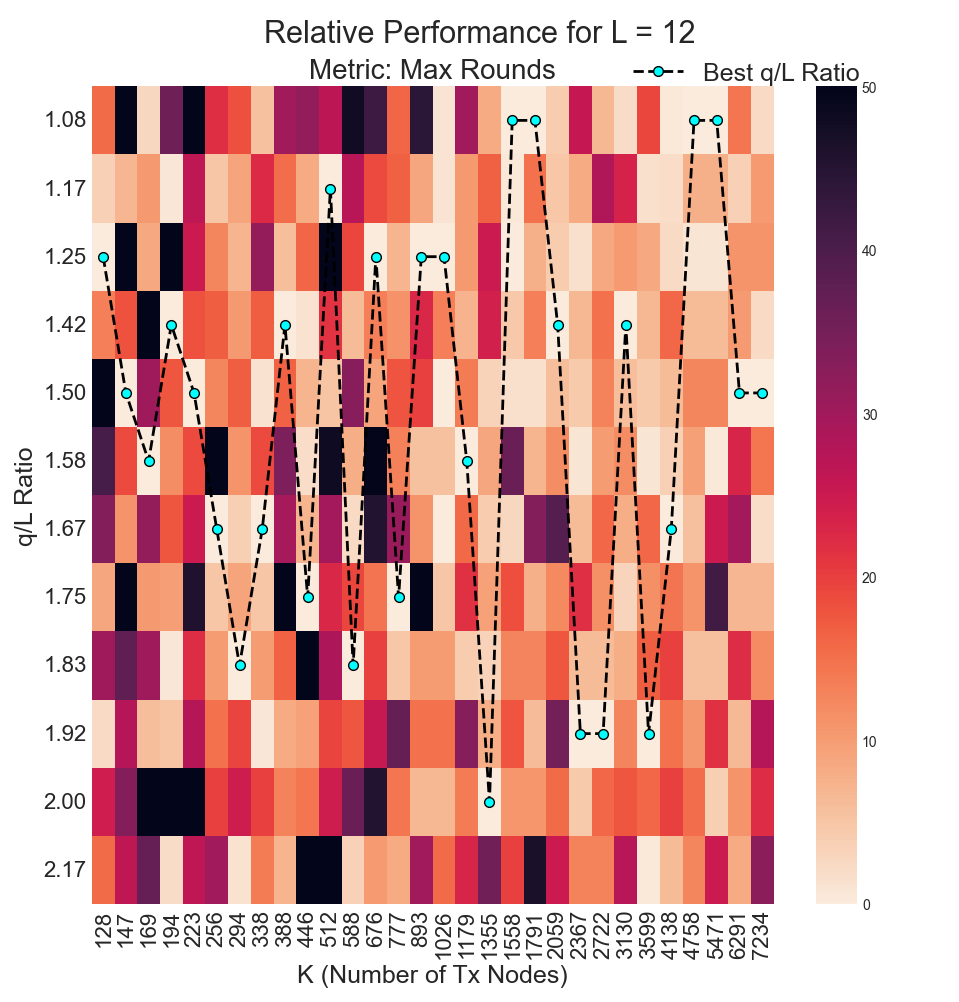}
    \caption{Relative degradation of the maximum network completion time versus \(K\) and \(q/L\). The optimum occurs near \(q/L\approx 1.6\).}
    \label{fig:heatmap_max}
\end{subfigure}
\caption{Heatmaps of PRISM performance as a function of network size \(K\) and the ratio \(q/L\)}
\label{fig:heatmap_combined}
\end{figure} 

\subsection{Baseline Algorithms and Adaptations}
\label{sec:baselines}

\begin{figure}[ht]
\centering
\includegraphics[width=\linewidth,height=0.85\textheight,keepaspectratio]{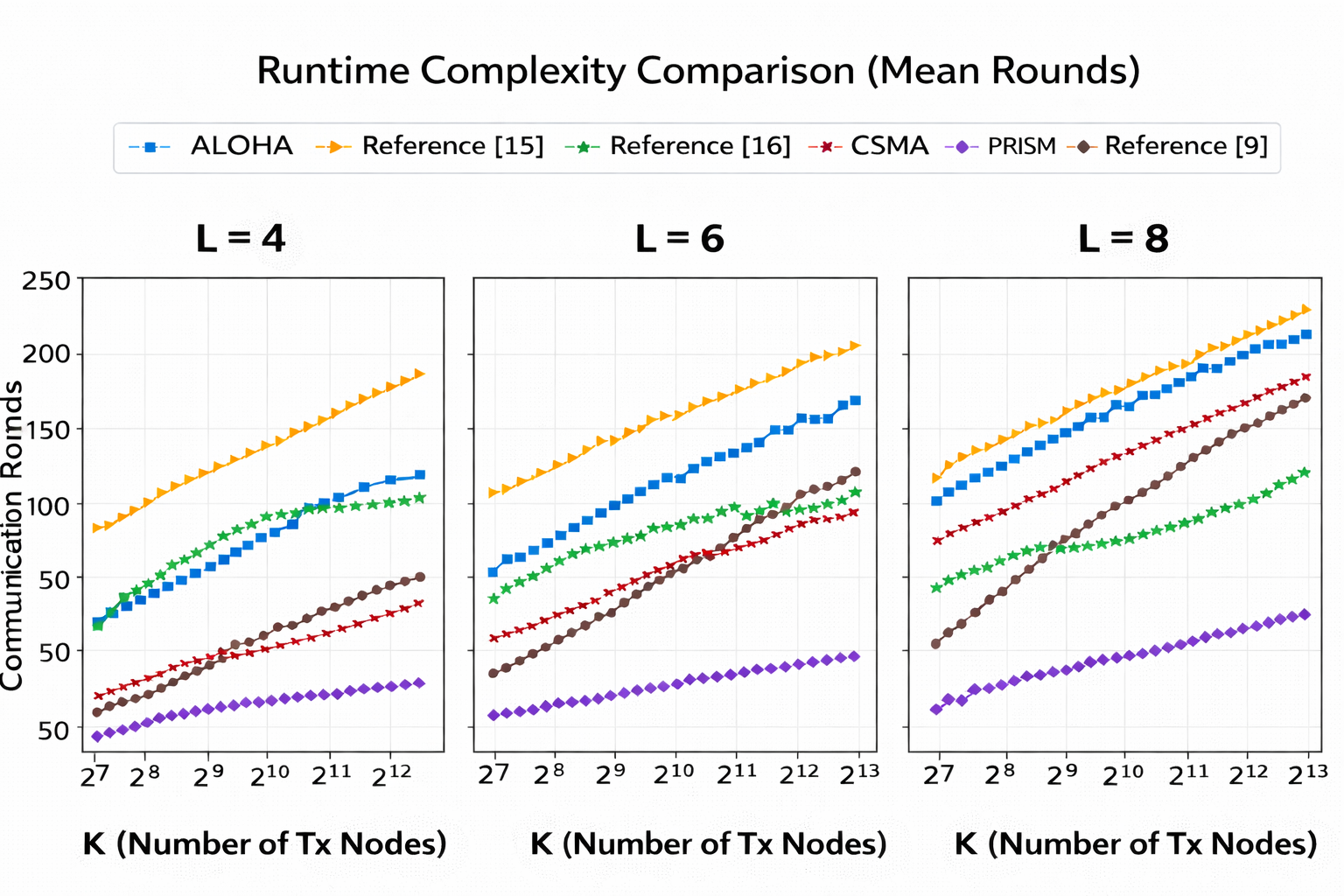}
\caption{Comparison of mean network completion time (discovery rounds) for topology discovery. PRISM outperforms \textsc{ALOHA}~\cite{ALOHA}, \textsc{CSMA}~\cite{CSMA},~\cite{Seyfi2021},~\cite{Choi16}, and~\cite{Chen23} across all tested \(K\) and \(L\).}
\label{fig:comparison}
\end{figure}

We compare PRISM with five baselines: \textsc{ALOHA}~\cite{ALOHA}, \textsc{CSMA}~\cite{CSMA}, the deterministic scheduling method of~\cite{Seyfi2021}, the block-design-based discovery scheme of~\cite{Choi16}, and the sparse OFDMA-based approach of~\cite{Chen23}. All methods are evaluated under the same slotted collision-channel abstraction, in which each round yields silence, a singleton, or a collision; only singleton observations reveal neighbors; and discovery terminates once every receiver identifies its complete neighborhood. For \textsc{ALOHA} and \textsc{CSMA}, each transmitter follows the standard random-access rule in the slotted setting. For~\cite{Seyfi2021}, we use the original deterministic cyclic schedule and map each scheduled transmission opportunity to one discovery round. For~\cite{Choi16}, we retain the deterministic on--off structure while adapting the original asynchronous design to synchronous round-based discovery. For~\cite{Chen23}, we preserve the sparse access graph and singleton-peeling dynamics while abstracting away the physical-layer processing. To ensure fairness, we disable auxiliary physical-layer features in the original methods, including asynchronous timing, energy-awareness mechanisms, power-domain separation, and successive interference cancellation. Thus, performance differences reflect discovery structure rather than PHY-layer enhancements. For each \((K,L)\) configuration, all methods are evaluated on the same \(200\) topology realizations. As shown in Figure~\ref{fig:comparison}, PRISM consistently requires fewer discovery rounds than all baselines, with the performance gap widening as \(K\) grows.

\section{Conclusion}

We presented PRISM, a deterministic and non-interactive framework for topology discovery in single-hop wireless networks under collision-channel constraints. PRISM uses modular arithmetic over multiplicative residue classes to construct fixed discovery schedules without feedback or probabilistic access. We showed that PRISM achieves complete discovery in \(O(L(1+\delta)\log K)\) rounds in expectation and in \(O(L^2\log K)\) rounds in the worst case. Simulations support these results, showing empirical scaling close to \(0.9\,L\log K\), with \(q/L\approx 1.2\) minimizing mean completion time and \(q/L\approx 1.4\)--\(1.6\) improving tail performance. We also showed that deterministic convergence can be certified efficiently through linear-time residue-alignment checks, making PRISM practical for settings requiring verifiable worst-case guarantees.

\bibliographystyle{IEEEtran}  
\bibliography{References}

\end{document}